\documentclass[sigconf,nonacm]{acmart}

\usepackage{xcolor}
\usepackage{alltt}
\usepackage[verbatim]{minted}
\definecolor{LightGray}{gray}{0.93}
\definecolor{codegreen}{rgb}{0,0.6,0}
\usepackage{listings}
\lstdefinestyle{mystyle}{
    backgroundcolor=\color{LightGray},
    basicstyle=\ttfamily,
    frame=lrtb,
    framerule=0pt,
    xleftmargin=0.5em
}

\setminted{
    frame=lines,
    framesep=2mm,
    baselinestretch=1.2,
    bgcolor=LightGray,
    fontsize=\footnotesize,
    linenos,
    breaklines=false
}

\AtBeginDocument{%
  }

\begin{document}

\title{DuckDB on xNVMe}

\author{Marius Ottosen}
\affiliation{%
  \institution{Department of Computer Science, University of Copenhagen}
  \country{Denmark}
}
\email{mpo@di.ku.dk}

\author{Magnus Keinicke Parlo}
\affiliation{%
  \institution{Department of Computer Science, University of Copenhagen}
  \country{Denmark}
}
\email{mfh144@alumni.ku.dk}

\author{Philippe Bonnet}
\affiliation{%
  \institution{Department of Computer Science, University of Copenhagen}
  \country{Denmark}
}
\email{bonnet@di.ku.dk}

\renewcommand{\shortauthors}{Ottosen et al.}

\begin{abstract}
DuckDB is designed for portability. It is also designed to run anywhere, and possibly in contexts where it can be specialized for performance, e.g., as a cloud service or on a smart device.
In this paper, we consider the way DuckDB interacts with local storage. Our long term research question is whether and how SSDs could be co-designed with DuckDB.
As a first step towards vertical integration of DuckDB and programmable SSDs, we consider whether and how DuckDB can access NVMe SSDs directly.
By default, DuckDB relies on the POSIX file interface. In contrast, we rely on the xNVMe library and explore how it can be leveraged in DuckDB. 
We leverage the block-based nature of the DuckDB buffer manager to bypass the synchronous POSIX I/O interface,
the file system and the block manager. Instead, we directly issue asynchronous I/Os against the SSD logical block address space. 
Our preliminary experimental study compares different ways to manage asynchronous I/Os atop xNVMe.
The speed-up we observe over the DuckDB baseline is significant, even for the simplest scan query over a  TPC-H table.
As expected, the speed-up increases with the scale factor, and the Linux NVMe passthru improves performance.
Future work includes a more thorough experimental study, a flexible solution that combines raw NVMe access and legacy POSIX file interface as well
the co-design of DuckDB and SSDs.
\end{abstract}

\maketitle

\section{Introduction}
DuckDB\footnote{\url{www.duckdb.org}}~\cite{Raasveldt2019-lp} is an in-process OLAP Database Management System,
designed for simplicity of use and portability. It is also desgined to be run anywhere~\cite{djikstra}, including
environments where vertical integration is possible, e.g., in the cloud or on embedded devices.
In this paper, we focus on how to leverage NVMe SSDs with significant but minimal changes to DuckDB.
 
DuckDB relies on \textit{single file storage} on the local file system, accessed through the synchronous POSIX I/O interface. 
Our first hypothesis is that  \textit{performance gain can be achieved by accessing NVMe SSDs directly and bypassing the file system.} 

To access NVMe SSDSs, we rely on \textbf{xNVMe}~\cite{Lund2020-hj,Lund2022-sx}.
xNVMe is a library for managing \textit{NVMe SSD Devices} explicitly and directly through a variety of I/O interfaces that can be
chosen dynamically at run-time.Our second hypothesis is that \textit{xNVMe provides abstractions that can be leveraged in DuckDB with minimal re-design}. 

This short paper, based on a MSc thesis\footnote{The thesis is titled \texttt{xNVMe on DuckDB} by Marius Ottosen and Magnus Keinicke Parlo, defended
at University of Copenhagen in June 2025. The manuscript is public and available from the Danish Royal Library~\url{https://soeg.kb.dk/discovery/search?vid=45KBDK_KGL:SPECIALER&lang=en}},
explains our design and reports on our experimental results.
In conclusion, we discuss potential directions for future work.

\section{Related Work}

The benefits of bypassing the file system and accessing NVMe SSDs directly have been demonstrated by modern OLTP database systems (e.g., Leanstore~\cite{Haas2020-rj,Haas2023-lw}, Umbra~\cite{Neumann2020-it} or Tigerbeetle\footnote{\url{https://tigerbeetle.com}}).
More generally, how to replace synchronous I/Os by asynchronous I/Os in a database system is an issue that has been studied for 20 years~\cite{Hall2005-db, Von_Merzljak2022-fu}.
In this project, we revisit these issues with DuckDB atop xNVMe.

Early work on xNVMe~\cite{Lund2022-sx} showed that some performance benefits can actually be seen by using xNVMe compared to using standard I/O Interfaces directly.
These improvements are not significantly high, but promising as they defy the expectation of higher latency - due to the overhead of the xNVMe API. Furthermore, when
adjusting variables like queue depth and block size it is shown that xNVMe provides near optimal results - without significant degradation in latency due to overhead.

A very similar idea to what we explore in this project was proposed in another xNVMe paper.\cite{xnvme-codesign} Here, the authors mention their interest in integrating
xNVMe with open-source DBMSs. One example mentioned of work in this direction is on the DBMS \textbf{RocksDB}\footnote{https://rocksdb.org/}, a persistent storage key-value store from Meta. 
A follow-up paper~\cite{cidr26} focuses on the integration of xNVMe with DuckDB, as an extension, via a specialized file system. This paper is not published at the time of
writing, but discussions with Vivek Shah point out that the focus of the paper is on the portability provided by xNVMe across a range of I/O interfaces. By contrast, our design is
not based on an extension but on specialization of the DuckDB storage manager atop xNVMe. 

How to leverage SSDs for off-core spilling during query processing using the POSIX interface is thoroughly documented~\cite{Kuiper2021-uy,Kuschewski2024-ty,Kuiper2023-pz}.
Whether and how these off-core algorithms benefit from direct integration with SSDs through xNVMe is most relevant and a topic for future work.

\section{Background}

\subsection{xNVMe}

When developing software, one I/O Interface may prove more useful or efficient than others, so naturally it would be a good idea to think critically about this choice and treat it as a
form of parameter for the development process and final product quality. However, as different I/O Interfaces typically will be available as some library with functionality,
it is rarely feasible for a developer to try multiple options - as it would simply be too involved to integrate all such libraries.

Furthermore, not all I/O libraries provide functionality for direct communication with (SSD) devices. Indeed, exploiting the NVMe Standard is not even an option for most setups today -
without looking into more low-level operations that are typically prone to implementation errors/vulnerabilities.

\begin{quote}
    \textit{"Today, picking an I/O interface is a difficult choice that has a deep impact on system design.
    Picking a single I/O interface is problematic because the entire system inherits its limitations in terms of
    portability, expressiveness or performance."}
    \cite{xnvme-presentation}
\end{quote}

These challenges are exactly what \texttt{xNVMe}\footnote{https://github.com/xnvme/xnvme} aims to tackle. As the name suggests, the main focus is on utilizing NVMe capabilities -
but actually, any device can be used with it.\cite{xnvme-codesign} However, not only does \texttt{xNVMe} allow for communication through the NVMe Standard, but also it introduces
an interface that fully abstracts the underlying I/O Interface - allowing for \textbf{I/O Independence}. Below we will introduce some of the relevant parts of the \texttt{xNVMe} design,
before going into more technical detail later in this report.
\cite{xnvme-presentation}

\subsubsection{I/O Independence}
\label{subsubsec:xnvme-io_independence}

Instead of finding, integrating and building an application with different I/O Interfaces, this choice can now be made at runtime, using \texttt{xNVMe}. It provides a typical
I/O Interface, allowing for issuance of read/write/etc. commands, but this issuance now include also some \textbf{I/O options} - where the underlying I/O Interface is set,
among othee configurations. Now, libraries, like io\_uring, libaio, aio, etc., can be exchanged at run-time, by setting a variable, as follows:
\begin{figure}[H]
\centering
\begin{verbatim}
struct xnvme opts;
if (cond) {
    opts = {.async = "libaio"}
} else {
    opts = {.async = "io_uring"};
}
\end{verbatim}

\caption{Example of deciding between either \texttt{libaio} or \texttt{io\_uring} as the underlying I/O Interface for \texttt{xNVMe}.}
\label{fig:code-xnvme_io_option}
\end{figure}

This is great, as it allows a developer to think of I/O Interfaces as just a parameter for development, rather than a larger integration part of the code.
One of the key aspects of xNVMe is that it is a convenient way to leverage I/O interface innovations without re-designing or re-implementing a data system.
It suffices that xNVMe supports an I/O interface feature for it to be available to a data system.
In the experimental section, we measure the impact of NVMe passthrough~\cite{Joshi2024-nd} on DuckDB performance.
This is not an issue that we consider in our design. We focus on xNVMe integration and just configure the underlying passthrough at run-time.

Lastly, it should be added that the \texttt{xNVMe} creators find the added overhead of such I/O Independence to have negligible effect on overall performance~\cite{xnvme-presentation}.

\subsubsection{Direct NVMe Device Access}

Another benefit of using \texttt{xNVMe} is the possibility of direct NVMe communication, which is actually the interesting part in terms of this project, and not so much the I/O Independence.

\begin{quote}
    \textit{"Leveraging such high-performance SSDs requires a streamlined storage stack that provides user-space software with (i) control over allocation and layout policies,
	(ii) control over I/O scheduling, and (iii) an I/O path without redundancies or missed optimization opportunities."}
    \cite{xnvme-presentation}
\end{quote}

All interaction with the \texttt{xNVMe} interface happens through a device (\texttt{xnvme\_dev}) object, which can be both a device and a file, depending on the path given
to the \texttt{xnvme\_dev\_open} function - this is something we will talk more about later. When a \textbf{command} is issued for a device, it typically carries some \textbf{meta information}
together with a \textbf{payload}. This is then transferred through the NVMe Standard as commands (from a command set) and DMA buffers. \texttt{xNVMe} allows for both synchronous and asynchronous
handling of commands. The difference is whether or not the submission/completion queue(s) are visible/accessible to the user.

To handle payloads, \texttt{xNVMe} also provides functionality to allocate DMA buffers while ensuring correct DMA buffer transfer - by for example ensuring correct byte alignment. Furthermore,
as presented above, the NVMe Standard allows for much configuration for device communication/use. When opening a device \texttt{xNVMe} can \textbf{automatically infer}
default options for a \textbf{device handle}, but as already introduced the user can manually overwrite these. Besides the run-time choice of I/O Interface, options can
also decide how the device should manage asynchronous activity, which command set to use, permission flags, and more. We will go into more detail about these options later.
\cite{xnvme-presentation}

\subsubsection{\texttt{libxnvme}}
\label{subsubsec:xnvme-libxnvme}

The centerbone of \texttt{xNVMe} is the library \texttt{libxnvme}, which provides all the interfaces and options described above, while managing the underlying interfaces too.
This is a good example of an \textbf{hourglass structure}, where many services can communicate with many interfaces, all through one central interface~\cite{Lund2022-sx}.

\begin{quote}
    \textit{"libxnvme makes it possible to (a) generate a command context by opening a device through a given I/O interface and creating queues for asynchronous I/Os and
	(b) allocate buffers for the payload in DMA transferrable memory if this is supported by the underlying I/O interface or else in virtual memory. libxnvme is thus a
	deep module whose interface defines five interrelated abstractions: devices, backends, queues, commands and buffers."}
    \cite{xnvme-presentation}
\end{quote}

The central component of \texttt{libxnvme}, making all the communication and translation possible, is the \textbf{Command Context} abstraction. Internally, \texttt{libxnvme}
uses the \texttt{xnvme\_cmd\_pass} interface, externally presenting interfaces that wrap this use, like e.g. the \texttt{xnvme\_nvm\_read} function. The Command Context carries
all the necessary information to pass between the internal components of \texttt{libxnvme}, the most relevant (which will be revisited later) are:
\begin{itemize}
\item First, the opened device for communication to go to and from, together with the options configuring how that communication should be managed. This will always be the actual
device in use, although it can potentially be "hidden" behind a file layer abstraction, depending on how the device is opened. An opened device can also be used to extract information
from the SSD, about e.g. \textbf{namespace} and \textbf{geometry}, that sets the boundaries within which the communication must happen.

\item Next, the \textbf{queue management} for managing commands to be/already processed. This is always part of a Command Context, but only directly visible and used by a program
if asynchronous management is explicitly used/controlled. Actually, \texttt{libxnvme} makes the handling of submission/completion queues (seen from the NVMe Standard) easier by
collecting them into one queue.

\item Lastly, the actual command and the completion result; this includes meta information about what command to execute and often the payload to be written or some buffer to read into.
\end{itemize}

\subsection{DuckDB}

In the context of this project, the most relevant part of DuckDB's design, is how DuckDB actually interact with the local file system - or, more generally, how it handles local storage.

Like other parts of the DuckDB design, this is done in a similar way to popular counterparts, like SQLite, through a \textbf{single-file storage format} - meaning, a Database is
stored in one single file. This can be either a temporary file (managed by the DBMS for a single session) or a persistent file (allowing for re-use of the same Database for multiple
sessions). Here, we will focus on how the local file system is used by DuckDB, which is fairly standard, but later we will look into alternative ways of managing such a Database file.

By storing a Database in a single file, somewhere on disk through the file system, we're able to easily keep track of the Database state like if it were any other file - allowing
for straightforward multi-version control, portability, backups, etc. - however, this of course imposes the standard limitations of a file system.

DuckDB's single file format is optimized for efficient scans and bulk manipulation, as this is the important use cases for OLAP work. A Database file is structured as a sequence of
blocks; a header block (3 x 4096 B) followed by data blocks (256 KB each). Furthermore, every block is always read/written in full between disk and memory.

The header of a Database file contains meta information for the health and state of the Database, e.g. storing the magic bytes ("DUCK"), a table catalog, and a list of free blocks. 
A catalog stores pointers to schemas, tables, and views - where tables is made up of the actual blocks storing data.

DuckDB manages the Database file by first updating data and then updating the pointers. The way DuckDB manages data blocks (and pointers between and inside of them) is through
\textbf{Adaptive Radix Tree (ART) Indexing}; a highly space efficient tree structure allowing for fast read operations, while also supporting efficient
insert/deletion~\cite{duckdb-design_details, duckdb-embedded_olap}.


ART is an example of a \textbf{Trie}, where each node functions as a value prefixing its subtrees: a node not only contains a value, but its position also plays an important
part of the context. The way DuckDB uses ART is to manage each block of a Database file, storing the content pointers~\cite{duckdb-art_blog, duckdb-art_paper}.


On top of this efficient indexing structure, DuckDB also supports lazy loading of nodes: despite managing blocks in full, only nodes that are actually relevant to the query being
processed are loaded into memory. To allow for this, DuckDB stores nodes of the ART structure by \textbf{Post-Order Traversal}, meaning children nodes must be stored "before" its parent inside a
block~\cite{duckdb-art_blog}.

\section{Analysis}

Where xNVMe presents mostly a flat set of APIs, with little inter-dependencies, DuckDB is a much larger code base with many layers and dependencies between
the different parts. We will here present the parts of the code we have found relevant, to different degrees, and how they interact to create the DuckDB control
flow, but there are certainly many parts of the code that we have not looked through - either due to irrelevance or time constraints.

\begin{figure*}[t]
    \centering
    \includegraphics[width=0.95\linewidth]{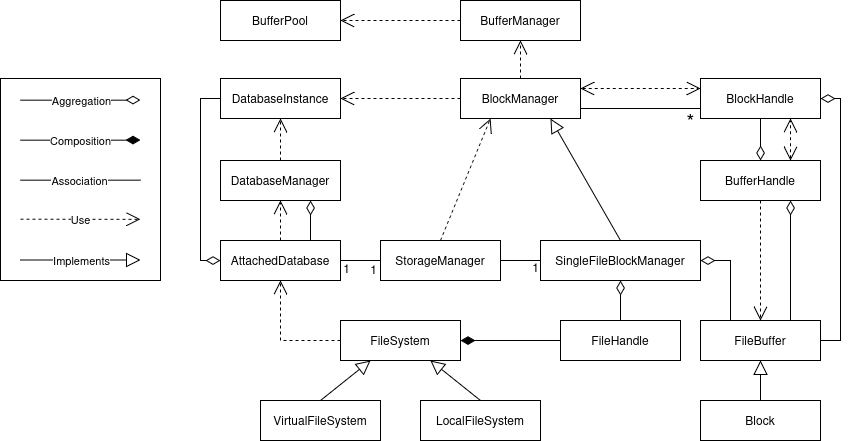}
    \caption{A cleaned up UML Class Diagram of the DuckDB design/structure, only including the relevant parts to our project. Due to the complexity, we have left out the inner definitions of each class.}
    \label{fig:analysis-duckdb_uml}
\end{figure*}

We present a \textbf{UML Class Diagram}, of the parts of DuckDB we have found most relevant, in \autoref{fig:analysis-duckdb_uml}. As this structure is quite complex,
we have decided to leave out details on definitions for the
individual classes, and instead we will go through these in this section. Also, this decision was made, as the classes contain a lot of functionality that is really not
directly relevant to our project, resulting in a very polluted diagram. Rather, we aim to here explain the interesting and relevant parts in text.

We will start by going in depth with the part of \autoref{fig:analysis-duckdb_uml} that is most closely related to our project, i.e., the lower right quadrant.
The main class, that our work has been centered around, is the \texttt{SingleFileBlockManager}. We will get back to this, but choose to work in a down-up fashion, starting with "underlying" classes.

First, we look at the File System setup, where DuckDB actually resembles the Linux standard very closely.
DuckDB implements, in code, abstractions for \texttt{VirtualFileSystem} and \texttt{LocalFileSystem}, which both implements the \texttt{FileSystem} Interface. These classes
implement all the traditional file system functionality, e.g., opening/closing files and reading/writing to them. Normally, the \texttt{LocalFileSystem} is used for most cases,
as DuckDB is embedded in existing programs, which can then both communicate with the local environment. DuckDB uses this class for many things, handling both Database
Files, Logs, and Temporary Files.

\begin{figure}[t]
    \centering
    \includegraphics[width=0.95\linewidth]{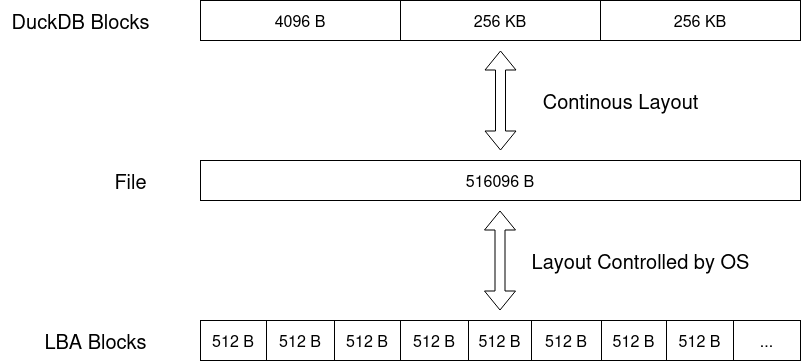}
    \caption{A high-level view of how DuckDB Blocks are managed as a single Database File, and finally stored somewhere on disk. In DuckDB, header blocks are 4096 B and data blocks are 256 KB. Logical block size on the SSD is 512B and
     the maximum data transfer size that we use for our mapping is 128KB.}
    \label{fig:analysis-duckdb_file_block_structure}
\end{figure}

As we are only focusing on the DuckDB Database Files, we look more closely at how these classes are used to manage these. The \texttt{LocalFileSystem} class is used to managing the
actual file that stores the Database, with the file stored in \texttt{SingleFileBlockManager} as a \texttt{FileHandle} object. The \texttt{SingleFileBlockManager} then manages the
structure and communication with the Database File through a \texttt{FileBuffer}. Before we go into more detail on this, we present in \autoref{fig:analysis-duckdb_file_block_structure}
 a high-level view of how DuckDB Blocks are managed as a single Database File, to finally be stored somewhere on disk by the OS.

\autoref{fig:analysis-duckdb_file_block_structure} shows how DuckDB manages a Database through multiple DuckDB Blocks - first some headers, followed by the actual data.
These files are written to the Database File in a continuous fashion, meaning the top level (DuckDB Blocks) in the figure would be layed out the same way in the mid level (File).
This is managed through Block IDs, that each is connected with some offset in the file. Then, when a DuckDB Block has been written to the Database File the read/write management of the file
from/to disk is managed by the OS.

To better understand how the control/data flow works between the levels presented in \autoref{fig:analysis-duckdb_file_block_structure}, we must look again at \autoref{fig:analysis-duckdb_uml}.
Here, it can be seen that \texttt{SingleFileBlockManager} implements the \texttt{BlockManager} class. Following the UML Class Diagram to the right, we see that \texttt{BlockManager} contains
multiple pointers to \texttt{BlockHandle} objects, that each contain a reference to a \texttt{FileBuffer}. This shows how DuckDB Blocks are managed through multiple buffers, and can therefore be
read/written concurrently from/to the Database File.

Managing this rather complex interplay between multiple classes is the \texttt{SingleFileBlockManager}, implementing the \texttt{BlockManager} class. Besides storing pointers to a \texttt{FileHandle}
and a \texttt{FileBuffer}, it also stores multiple \texttt{BlockHandle} pointers. Furthermore, to manage these DuckDB Blocks, the \texttt{SingleFileBlockManager} class maintains meta information on
these blocks. This includes a set of blocks that can be written to freely (\texttt{free\_list}), a set of blocks that have been modified and should be moved back to being free (\texttt{modified\_blocks}),
the maximum block id currently managed (\texttt{max\_block}), and finally a lock for managing the blocks of a file (\texttt{block\_lock}).

Finally, we take a look at the less relevant parts of \autoref{fig:analysis-duckdb_uml}, focusing on the upper-left quadrant. This part shows the more high-level DuckDB design for managing a database -
or multiple attached databases.

The most relevant part of these classes is the \texttt{StorageManager}, which does not directly manage file buffers, but as part of its functionality will open a database (new or existing) and
manage this through a \texttt{SingleFileBlockManager} object. This is where a specified path (by the user) is used for determining the type and state of a DuckDB Database, as well as the
Logging connected with that Database.

The \texttt{StorageManager} class is closely bound with the\\
\texttt{AttachedDatabase}, which further is closely related to the\\
\texttt{DatabaseInstance} class. Finally, most of this is
all managed as part of the top-level class \texttt{DatabaseManager}.

Another area of DuckDB, that's important to understand is concurrency. DuckDB attempts to divide the workload onto threads, for concurrent execution. To do that, DuckDB splits queries
into manageable tasks, which threads can pick up and execute. Results are then combined when all tasks for a specific operator are done, and the next operator of a query can be executed
via new tasks, using the results of the previous operator.

\begin{figure}[H]
    \centering
    \includegraphics[width=0.95\linewidth]{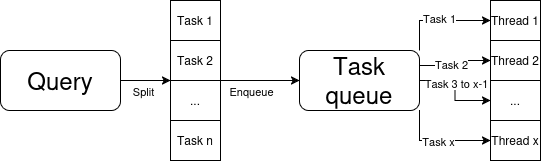}
    \caption{Queries are divided into tasks for the threads to execute concurrently. They are enqueued via a task scheduler, and they are picked up by the threads by popping from the queue in a
    FIFO manner. A query can consist of multiple operators. In that case, it is the specific operators that are split into tasks. This step is not shown in this figure.}
    \label{fig:analysis-duckdb_task_overview}
\end{figure}

This design begs the question: What if two tasks have to operate on the same DuckDB block, e.g. two tasks have to write different data to the same DuckDB block? The answer
is that DuckDB keeps track of dependencies between tasks. Using these dependencies, DuckDB will not enqueue a task until all dependencies for that task are resolved.
Since reads and writes in DuckDB are synchronous, correctness is guaranteed. The consequence of this analysis will be discussed later.

An important note on this single file structure is that this ART structure, and the content of a block, is managed in-memory, and blocks themselves are then read/written
to their designated space in the Database file on disk. We take advantage of this strong framework in our work, as we will be focusing specifically on
moving data (as blocks) between disk and memory. The assumption is that there is no need for a file system as ART provides the indexes that DuckDB needs
and that the DuckDB single file can be replaced by a statically allocated SSD partition. Here we consider that dynamic file resizing is neither necessary nor desirable. 
Our approach only makes sense if the underlying SSDs are dedicated to the database.

\section{Design and Implementation}

\begin{figure}[H]
    \centering
    \includegraphics[width=0.95\linewidth]{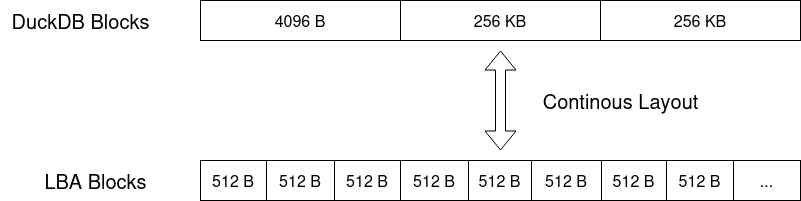}
    \caption{Illustration of our second design for modifying DuckDB, cutting out the File Layer from \autoref{fig:analysis-duckdb_file_block_structure}.}
    \label{fig:design-design2}
\end{figure}

We do not modify the DuckDB block structure, as presented in \autoref{fig:analysis-duckdb_uml}. We simply remove the file abstraction
and directly map DuckDB blocks onto the logical block address (LBA) space exposed by the SSD. The key insight is that the file offset stored by the DuckDB block
can trivially be mapped onto an offset in the SSD LBA space~\autoref{fig:design-design2}.

More specifically, we modify the \texttt{FileBuffer}, as this is where read/write functionality is found for managing Database Blocks. These functions use the \texttt{FileHandle},
as explained earlier, but we instead plan to substitute this with the xNVMe APIs for manipulating an NVMe device directly. 
Beside just modifying the \texttt{FileBuffer}, we  also need to modify the  \texttt{SingleFileBlockManager} slightly. \\
The \texttt{SingleFileBlockManager} currently stores both a pointer to the \texttt{FileHandle} and \texttt{FileBuffer}, but should instead store an\\
\texttt{xnvme\_dev} object together with the \texttt{FileBuffer}, as we're replacing the File with the raw device.\\

For mapping DuckDB Blocks to the device LBA space, there are some considerations to be made. Normally, when writing DuckDB Blocks to a file, by an offset
from the file pointer, the OS will be responsible for placing parts of the file on actual disk. However, we must take care of this; mapping a file offset to an LBA offset.
The trivial translation is simply to divide a file offset by the LBA block size, and submitting LBA block size commands to over NVMe. However, through talking with Simon Lund,
the creator of xNVMe, it would be more efficient to submit Maximum Data Transfer Size (MDTS) size commands to the device. IN our system, MDTS is 128 KB. Hence, we design this mapping to be file offset
divided by mdts size of the device\footnote{By using a device to store data, the same way we would a file, we will be using a logical address space strongly
biased to the start of the device. However, making sure the physical addresses are distributed over time across different cells on the device is the responsibility of the
FTL.}.
 

There are two different ways to go about implementing this design, using NVMe through xNVMe. 
\begin{enumerate}
    \item \textbf{Synchronous I/O}: This maintains the correctness guarantees that DuckDB already has, by blocking until I/O commands have completed.
    \item \textbf{Asynchronous I/O}: This approach uses submission and completion queues to handle I/O asynchronously.
    This approach is expected to be faster than the synchronous approach, but it also violates the guarantees of correctness that DuckDB gives in the
    original implementation. This is illustrated in \autoref{fig:design-duckdb_dependent_tasks} below.
\end{enumerate}

\begin{figure}[H]
        \centering
        \includegraphics[width=0.45\linewidth]{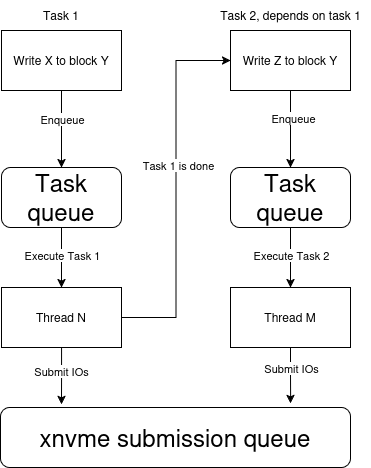}
        \caption{Task 2 depends on Task 1, so it is only enqueued when Task 1 has finished. Both Task 1 and Task 2 submit NVMe commands to the submission queue.
        Even though Task 1 is marked as done, we cannot be sure that the I/O commands that it submitted has completed. We end up in a race condition where the IO
        submissions of Task 2 might be completed before the I/O submissions of Task 1, which violates the correctness guarantee.}
        \label{fig:design-duckdb_dependent_tasks}
    \end{figure}

When working asynchronously, there is a risk of race conditions for the scheduling of tasks. We will go into more detail on our considerations
for this and how we can mitigate these risks in our implementation below. The synchronous design and implementation is described in the MSc thesis.
In the rest of this report, we focus on asynchronous I/Os.

\subsection{Asynchronous I/O with one NVMe Queue}

When presenting our design for submitting commands, above, we have mentioned context queue(s) multiple times - something that actually plays a much bigger role in our
design, than what we were initially expecting.

While we do not implement an actual \texttt{NvmeHandle} for this design, we do implement two classes; \texttt{QueuePool} and \texttt{QueueWrapper}. These classes are used to
support the asynchronous capabilities of the NVMe Standard - working with submission/completion queues. The \texttt{QueuePool}
is not needed when only one queue is present, so we present that later.

First, the \texttt{QueueWrapper} class, as the name suggests, wraps an individual context queue. With xNVMe, instead of managing both a submission and a completion queue,
these are combined into just one queue, which makes it easier to keep track of pointers. However, due to combining these two, we need some way to keep track of the ratio between
submissions and completions, which is done through a flat struct object connected with each \texttt{QueueWrapper}. Also, making sure this struct is kept up to date is a callback
function, which is triggered when a command is completed.

For asynchronous I/O, we have some challenges that we have to consider. The first challenge was presented in \autoref{fig:design-duckdb_dependent_tasks}. Since NVMe queues are not
thread safe, we have to make sure that they are not accessed by two threads concurrently. With only one queue, each thread needs to submit reads and writes to the same queue, so we
need a locking mechanism to avoid race conditions. To do that, we added a mutex to the \texttt{QueueWrapper} that each thread needs to lock, in order to submit commands, as well as
 draining the queue.

Normally, when writing to a file, data is first placed in standard memory buffers - or the opposite when reading from a file into such buffers. However, to efficiently transfer data to DMA buffers, xNVMe provides their own \texttt{xnvme\_buf} type of buffers, that should be used. Our initial way to work around this was to simply copy relevant data between allocated memory buffers and xnvme buffers before reading/writing. However, we found that these copies induced too much overhead.
To improve our solution, we intended to use the \texttt{header\_buffer} object (of \texttt{SingleFileBlockManager}) together with an \texttt{xnvme\_dev}. However, only this only concerned header blocks, and hence data blocks did not benefit. Instead, we
got back to the \texttt{FileBuffer}; used throughout the code, as shown in Figure 2.
To provide and use xnvme buffers in the FileBuffer class, we ran into the issue that an xnvme device must be opened and present.
Our solution at the time only provided an xnvme device as part of the \texttt{SingleFileBlockManager},
why we needed to move this higher up in the code hierarchy. Our solution was to open
a device, when a database is opened/created at the top level.

\begin{listing}[H]
\begin{minted}{cpp}
int QueueWrapper::SubmitRead(xnvme_dev *dev, uint64_t lba_location,
                             uint16_t amount, data_ptr_t payload) {
	mtx.lock();
	struct xnvme_cmd_ctx *ctx = xnvme_queue_get_cmd_ctx(queue);

submit:
	int err = xnvme_nvm_read(ctx, xnvme_dev_get_nsid(dev),
	                         lba_location, amount,
	                         payload, nullptr);
	switch (err) {
	case 0:
		args.submitted++;
		break;
	case -EBUSY:
	case -EAGAIN:
		this->Poke();
		goto submit;
	default:
		xnvme_cli_perr("xnvme_nvm_read()", errno);
		xnvme_queue_put_cmd_ctx(queue, ctx);
		break;
	}

	mtx.unlock();
	return err;
}
\end{minted}
\caption{The queue wrapper locks the mutex before anything is done to the queue.}
\label{lst:implementation-queuewrapper_submitread_one_queue}
\end{listing}

\begin{listing}[H]
\begin{minted}{cpp}
void FileBuffer::Read(xnvme_dev *dev, uint64_t location,
                      QueueWrapper &queue) {
  D_ASSERT(type != FileBufferType::TINY_BUFFER);

  // extract meta data from device
  auto geo = xnvme_dev_get_geo(dev);
  auto lba_size = geo->nbytes;
  auto lba_location = location / lba_size;
  auto mdts_size = geo->mdts_nbytes;
  auto lbas_pr_mdts = mdts_size / lba_size;
  uint64_t submissions =
       1 + ((internal_size - 1) / mdts_size);
  for (uint64_t i = 0; i < submissions; i++) {
    auto offset = i * mdts_size;
    auto *payload = internal_buffer + offset;
    auto lbas = internal_size - offset >= mdts_size ?
      lbas_pr_mdts : (internal_size - offset) / lba_size;
    queue.SubmitRead(dev, lba_location + i * lbas_pr_mdts,
  	                 (uint16_t)lbas - 1, payload);
  }
  queue.Drain();
}
\end{minted}
\caption{The filebuffer submits read requests to the queue through the queue wrapper API in \autoref{lst:implementation-queuewrapper_submitread_one_queue}. The Write function is analogous.}
\label{lst:implementation-filebuffer_read_one_queue}
\end{listing}

\subsection{Asynchronous I/O with multiple NVMe Queues}

\begin{listing}[t]
\begin{minted}{cpp}
void FileBuffer::Read(xnvme_dev *dev, uint64_t location,
                      QueuePool &qpool) {
  D_ASSERT(type != FileBufferType::TINY_BUFFER);

  // extract meta data from device
  auto geo = xnvme_dev_get_geo(dev);
  auto lba_size = geo->nbytes;
  auto lba_location = location / lba_size;
  auto mdts_size = geo->mdts_nbytes;
  auto lbas_pr_mdts = mdts_size / lba_size;
  uint64_t submissions =
    1 + ((internal_size - 1) / mdts_size);

  for (uint64_t i = 0; i < submissions; i++) {
    auto offset = i * mdts_size;
    auto *payload = internal_buffer + offset;
    auto lbas = internal_size - offset >= mdts_size ?
    lbas_pr_mdts : (internal_size-offset)/ lba_size;

    qpool.SubmitRead(dev,
                     lba_location + i * lbas_pr_mdts,
                     (uint16_t)lbas - 1, payload);
  }
}
\end{minted}
\caption{The asynchronous implementation of the \texttt{Read} function, with multiple NVMe queues combined in a \texttt{Queue Pool}. The write function is analogous}
\label{lst:implementation-filebuffer_read_design2}
\end{listing}

The \texttt{QueuePool} class\footnote{\url{https://github.com/Ma-Master-DK/duckdb/tree/nvme\_mdtsblock}} manages the support for maintaining multiple queues at once.
Eeach queue can hold thousands of entries, but an NVMe Device can furthermore handle many queues also -
generally, the recommendation is to have one queue per available thread on the system.

To show how e.g. a read to an NVMe Device might look, by submitting commands through a \texttt{QueuePool}, we present our implementation for \texttt{SubmitRead} for
both the QueuePool and QueueWrapper class in Listings \autoref{lst:implementation-queuepool_read} and \autoref{lst:implementation-queuewrapper_read}. Here, we see how we can submit a \texttt{Read} command to
a queue pool, which will then loop over the contained queues, looking for the first available (meaning, not full or already in use) queue to submit a command to. By this design,
we ensure that individual queues manages their own lock (to support integral actions like draining and terminating a queue), as well as support for distributing MDTS Blocks across
multiple queues, if necessary. However, it should be noted that the load on the queue pool will usually be biased towards the "first" queues that is looped through for each submitted
command. We have not implemented a more complex submission system, for more uniform submission across queues. Compared with the single queue approach, this time threads can actually
submit IOs concurrently, but to different queues. We expect to see better performance for that reason.

\begin{listing}[t]
\begin{minted}{cpp}
void QueuePool::SubmitRead(xnvme_dev *dev, uint64_t lba_location,
                           uint16_t amount, data_ptr_t payload) {
  while (true) {
    for (auto &qwrap_ptr : queues) {
      if (qwrap_ptr->SubmitRead(dev, lba_location,
          amount, payload) == 0) {
        return;
      }
    }
  }
}
\end{minted}
\caption{Our Design 2 implementation of \texttt{SubmitRead} for \texttt{QueuePool}.}
\label{lst:implementation-queuepool_read}
\end{listing}

\begin{listing}[t]
\begin{minted}{cpp}
int QueueWrapper::SubmitRead(xnvme_dev *dev, uint64_t lba_location,
                             uint16_t amount, data_ptr_t payload) {
  if (queue) {
    if (TryLock()) {
      struct xnvme_cmd_ctx *ctx =
                  xnvme_queue_get_cmd_ctx(queue);

      int err = xnvme_nvm_read(ctx,
                   xnvme_dev_get_nsid(dev), lba_location,
                   amount, payload, nullptr);
      if (err == 0) {
        args.submitted++;
        args.inflight++;
      }

      mtx.unlock();
      return err;
    } else {
      return -1;
    }
  }

  return -1;
}
\end{minted}
\caption{Our Design 2 implementation of \texttt{SubmitRead} for \texttt{QueueWrapper}.}
\label{lst:implementation-queuewrapper_read}
\end{listing}

An important note is that Listing \autoref{lst:implementation-filebuffer_read_design2} shows a lack of queue draining. Surprisingly, this still works, even though the challenge
presented in \autoref{fig:design-duckdb_dependent_tasks} is not resolved, thus the guarantee of correctness is not restored, however tests show that this solution actually
does produce the correct results. We suspect that IO completions happen fast enough so that whatever is submitted first will be completed first, but this is not a guarantee
that our code provides. An easy fix is to drain the queue we submitted to, right after we are done submitting the DuckDB block.

\subsection{Asynchronous I/O with thread owned NVMe Queues}
\label{subse:implementation-nvme_thread_queues}

When asking \textit{Simon Lund}, the creator of xnvme, how many queues we should use, he answered that the best approach is to assign a queue to each thread. This
approach eliminates the need to manage queue access, which avoids mutex locking. Instead, when the threads are initialized, we also initialize a queue that is owned and managed
by that thread. The thread will then submit commands to the queue and drain the queue after a DuckDB block has been submitted.
This approach automatically restores the correctness guarantee that DuckDB provides, as a task is only completed when the block IO commands have been completed, as the queue drain ensures.

\begin{listing}[t]
\begin{minted}{cpp}
namespace duckdb {
thread_local xnvme_queue *queue_ptr = nullptr;
}
\end{minted}
\caption{We define a header with a global thread local variable, which we can use anywhere to access the queue owned by the thread.}
\label{lst:implementation-global}
\end{listing}

\begin{listing}[t]
\begin{minted}{cpp}
static void ThreadExecuteTasks(TaskScheduler *scheduler,
                               atomic<bool> *marker,
                               xnvme_dev *dev) {
	xnvme_queue_init(dev, (uint16_t)64, 0, &queue_ptr);
	xnvme_queue_set_cb(queue_ptr, cb_fn, nullptr);
	scheduler->ExecuteForever(marker);
}
\end{minted}
\caption{The function that the worker threads are running. Before entering the worker thread loop, each thread will initialize its own queue and set the global
variable from \autoref{lst:implementation-global}. The queue is terminated in the end of the \texttt{ExecuteForever} function.}
\end{listing}

\begin{listing}[t]
\begin{minted}{cpp}
void FileBuffer::Read(xnvme_dev *dev, uint64_t location,
                      const xnvme_geo *geo) {
  D_ASSERT(type != FileBufferType::TINY_BUFFER);

  auto lbas_pr_mdts = geo->mdts_nbytes / geo->lba_nbytes;
  auto lba_location = location / geo->lba_nbytes;
  uint64_t submissions = 1 + ((internal_size - 1) / geo->mdts_nbytes);
  for (uint64_t i = 0; i < submissions; i++) {
    auto offset = i * geo->mdts_nbytes;
    auto *payload = internal_buffer + offset;
    auto lbas =
        internal_size - offset >= geo->mdts_nbytes ?
        lbas_pr_mdts : (internal_size - offset) / geo->lba_nbytes;
    struct xnvme_cmd_ctx *ctx = xnvme_queue_get_cmd_ctx(queue_ptr);
  submit:
    int err = xnvme_nvm_read(ctx, xnvme_dev_get_nsid(dev),
  	                         lba_location + i * lbas_pr_mdts,
  	                         (uint16_t)lbas - 1,
  	                         payload, nullptr);
  	switch (err) {
  	case 0:
  	  break;
  	case -EBUSY:
  	case -EAGAIN:
  	  xnvme_queue_poke(queue_ptr, 0);
  	  goto submit;
  	default:
  	  xnvme_cli_perr("xnvme_nvm_read()", err);
  	  xnvme_queue_put_cmd_ctx(queue_ptr, ctx);
  	  break;
    }
  }
  int ret = xnvme_queue_drain(queue_ptr);
  if (ret < 0) {
    xnvme_cli_perr("xnvme_queue_drain()", ret);
  }
}
\end{minted}
\caption{The filebuffer uses the thread\_local queue pointer from \autoref{lst:implementation-global} to submit IO commands to the queue owned by that thread.}
\label{lst:implementation-filebuffer_async_thread_queue}
\end{listing}

This solution\footnote{\url{https://github.com/Ma-Master-DK/duckdb/tree/nvme\_thread\_queues}} does not use the queue wrapper and queue pool abstractions, but the submission logic is directly calculated and passed to the xnvme API
in the file buffer. This solution removes the overhead created in the previous multi-queue async approach, but still allows the same level of concurrency
that the standard DuckDB program has. Additionally, using the queue, we allow aynchronous reading and writing of each individual DuckDB block, so we
expect that this solution should be the best performing solution of our implementations.

\section{Experimental Framework}

\subsection{System}
We run all experiments on the same system, presented below in \autoref{tab:experiment_environment}.

\begin{table}[t]
    \centering
    \begin{tabular}{|l|l|} \hline
        \textbf{Hardware} & \textbf{Model} \\\hline 
        Board & Framework FRANMZCP07 A7 \\\hline
        CPU & AMD Ryzen™ 7 7840HS, 8 Cores @ 3.800GHz \\\hline
        GPU & Radeon™ 780M Graphics \\\hline
        Memory & DDR5-5600 - 32GB \\\hline
        SSD & WD\_BLACK SN770 NVMe M.2 2280 - 500GB \\\hline \hline
        
        \textbf{Software} & \textbf{Model} \\\hline 
        OS & Debian GNU/Linux 12 (bookworm) x86\_64 \\\hline
        Kernel & 6.1.0-37-amd64 \\\hline
        DuckDB & v1.2.1 \\\hline
        xNVMe & v0.7.5 \\\hline
        gcc & v12.2.0 \\\hline
        clang & v14.0.6 \\\hline
        Python & v3.11.2 \\\hline
    \end{tabular}
    \caption{An overview of the system all our experiments will be conducted on.}
    \label{tab:experiment_environment}
\end{table}

DuckDB ships with an extensive test setup, allowing for thorough testing and benchmarking through a simple CLI and SQL interface.
As default, when building DuckDB from source, a wide test suite of functionality is available through a set of \texttt{C++} files.
This includes testing persistence, locking, optimization, serialization, leaks, logging, etc. We have tried to run different parts of this test suite,
to get familiar with the setup and interface, but found it incompatible with our modifications.
This is unfortunate, as DuckDB also provides a setup for benchmarking performance, which is really what we are interested in.
The way it works, however, is through multiple temporary POSIX files, targeted directly with different sets of queries.
In contrast, we only consider data written to disk through the buffer manager.
To use these tests, we would either need an additional step that imports data into DuckDB via the buffer manager, or more interestingly
we should extend our design with the possibility to access an existing a given file (exposed through a kernel module as collection of LBAs). 

We compare the standard implementation of DuckDB atop POSIX, with several variants of our integration of xNVMe: using files through xNVMe,
using synchronous I/Os, using asynchronous I/Os with a single queue, using asynchronous I/Os with a queue pool and
using asynchronous I/Os with thread queues. We also turn NVMe passthrough on and off within xNVMe without any modification of DuckDB.

\subsection{Sanity Check}

DuckDB is an OLAP RDBMS, so it excels at reading large amounts of data. Our experiments are a sanity check for the design of our system.
We focus on the best case for the POSIX interface, i.e., sequential scan.
With a sequential scan, the block layer can coalesce requests and prefetch blocks. In contrast, we access DuckDB blocks asynchronously.
The goal is that our design should not lag behind the POSIX interface. We measure query latency for a scan query on the TPC-H customer table
at various scales.

We design two experiments:
\begin{enumerate}
    \item \textbf{Passthrough:} First, we pick one of our asynchronous implementations to test the effect of using \texttt{io\_uring} with
    and without passthrough. We choose our thread-owned multi-queue implementation, as we argue this is our best implementation and likely best performing modification.
    This we translate to the null hypothesis: \textit{io\_uring without passthrough is at least as fast as with passthrough}.
    \item \textbf{Sanity check:} Here, we compare all our implementations against the standard DuckDB. From the first experiment we know whether passthrough is faster than standard \texttt{io\_uring}, so we choose the best performing version for our asynchronous implementations.\\
    This we translate to the null hypothesis: \textit{the mean query execution time for the standard DuckDB implementation is not slower than that of any of our implementations}.
\end{enumerate}

The workload is built around TPC-H benchmarking datasets. This is because DuckDB has a convenience query that generates these datasets with given size, called a \textbf{Scale Factor},
and associating TPC-H queries with associated results, so we can make sure that the database is correct after generation\footnote{Details on these DuckDB capabilities are provided
here:~\url{https://duckdb.org/docs/stable/core\_extensions/tpch.html}}. 

DuckDB also has another convenient query, \texttt{EXPLAIN ANALYZE}, which we can prepend to almost any query. This query will print the query plan and how long it took for
 each operator to finish execution. A total time is also printed, which is the wall clock time it took to execute the entire query. This will be the basis for our test results
as it allows us to time just the execution of an SQL statement, without including overhead like opening a Database file.\\

For our experiments, we use various Scale Factors up to 100. This Scale Factor provides a database that is 26 GB in size. It uses 71 GB of RAM to generate the dataset, and also takes
around 20 minutes. This database size is big enough to show some significant differences between the implementations, but small enough so that it is feasible to generate within
time and resource constraints.\\

Before presenting the results, we quickly summarize the expectations we identified from the implementation details.
\begin{itemize}
    \item \textbf{NVMe File:} We expect the file system based xNVMe implementation to be comparable to the standard DuckDB implementation. That is, we expect very similar performance between the two.
    \item \textbf{NVMe Sync:} The synchronous version communicates with the device through OS NVMe driver IOCTLs. This also bypasses a lot of layers, thus we will expect this to be faster than the standard DuckDB version, however since it uses sync I/O, we do not expect this to be the fastest implementation.
    \item \textbf{NVME Async, Single Queue:} This version uses only a single queue, thus threads cannot submit reads concurrently. For that reason, we expect this solution to be slow, since losing the threading concurrency will probably decrease performance more than the increase in disk concurrency within the single block reads.
    \item \textbf{NVMe Async, Queue Pool:} This version uses multiple queues, which fixes the problem from the previous implementation. Since there is still a lot of locking involved to avoid multiple threads accessing the same queue, we expect to have some overhead with this solution. We expect the solution to be faster than the standard DuckDB implementation, and the second fastest of our implementations.
    \item \textbf{NVMe Async, Threaded Queues:} This implementation avoids all the locking issues from the above implementation. For that reason, we expect this solution to be the absolute fastest.
\end{itemize}

We also expect that \texttt{io\_uring} is faster with passthrough than without.

\subsection{Statistical Confidence}

To confidently argue how our results compare, and whether we see a significant difference, to draw conclusions, we employ statistical \textbf{paired t-tests}\footnote{\url{https://www.statstutor.ac.uk/resources/uploaded/paired-t-test}}. 
From our p-tests, we get a p-value that we can use to argue what the probability is of seeing our results (or something more extreme/significant) if we assume the null hypothesis
is true. Generally, a p-value below 0.05 is considered sufficient to argue that results are significant. However, we will test against a p-value threshold of 0.01 - this will allow
us to be more confident in drawing conclusions. Complimentary to this, a p-value over 0.99 or higher would indicate the assumption of the null hypothesis might indeed be probable to be true.

To complement the above, we will also consider {\bf standard error} in our experiments. When testing for standard error, we're asking the question in what range we expect to observe the
mean of data points in the majority (68\%) of our experiments. As the sample size grows, the standard error range will grow if the data points have a high degree of noise. Hence, a small
standard error is desirable, as it would indicate that the data points show a low degree of noise, and therefore better grounds for arguing for the trustworthiness of our results.

\section{Experimental Results}

\subsection{Experiment 1: Passthrough}
\label{subsubsec:experiment-1}

\begin{figure}[H]
    \centering
    \includegraphics[width=\linewidth]{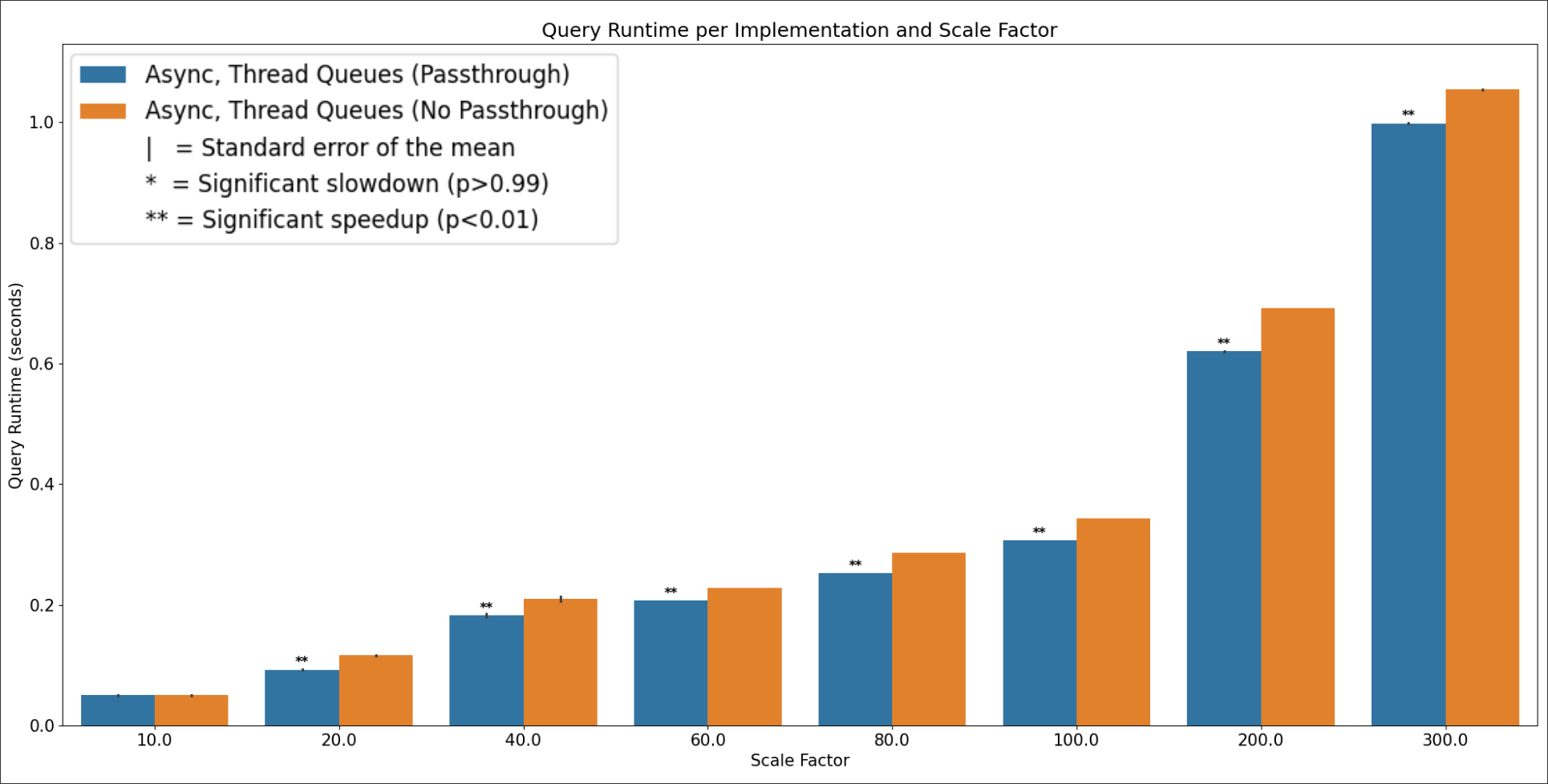}
    \caption{Bar plot of results from comparing Read Latencies with/without passthrough, we use our implementation with dedicated NVMe Queues per Thread.}
    \label{fig:results-passthrough_compare}
\end{figure}

In line with our expectations, we see that \texttt{io\_uring} is faster with passthrough than without, especially for larger scale factors. We do see some fluctuations, which
could posssibly be reduced with a larger amount of reads going into the calculation of the average read latency shown in \autoref{fig:results-passthrough_compare}.

This is a positive result, as it aligns with the theory of less latency for more direct communication. By bypassing layers in the OS, and talking directly with the NVMe driver,
we can optimize communication with a device, and ultimately lower latency.

Another relevant observation is that the latency increases substantially as the Scale Factor increases. This is also expected behavior, as we expect higer latency when processing more data.

We use our asynchronous implementations with passthrough for the following experiment.

\subsection{Experiment 2: Sanity Check}

\begin{figure}[H]
    \centering
    \includegraphics[width=\linewidth]{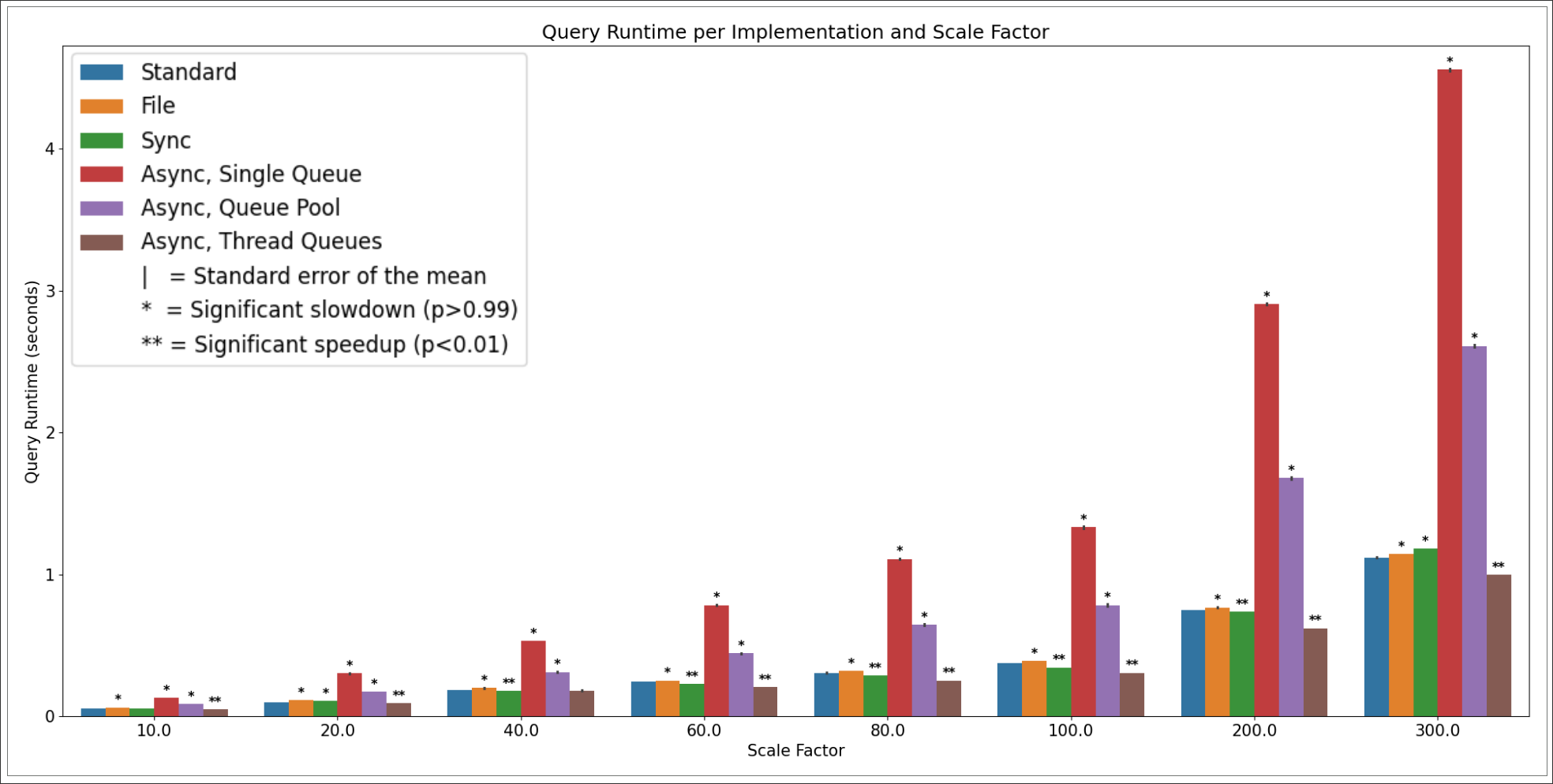}
    \caption{Bar plot of results from comparing Read Latencies between our implementations, against the Standard DuckDB implementation. All our async implementations here use passthrough.}
    \label{fig:results_implementation_compare}
\end{figure}

\begin{itemize}
    \item \textbf{NVMe File:} For the xNVMe file system implementation, we see that it is indeed very similar in performance to the standard DuckDB version. Interestingly,
    we see a slight trend of slower performance for smaller scale factors, while the largest scale factor reveal better performance. We speculate that larger datasets might
    reveal an even bigger difference.
    \item \textbf{NVMe Sync:} This implementation is fast, as we also expected. It is in fact the fastest implementation for a few of the scale factors, and does challenge
    the implementation we expected to be the fastest.
    \item \textbf{NVMe Async, Single Queue:} As expected, this implementation is slow, and we see that the loss of concurrency between DuckDB blocks is detrimental to performance.
    \item \textbf{NVMe Async, Queue Pool:} This solution is faster than the single queue, but it is still the second slowest implementation, which we did not expect. We postulate
    that the overhead of locking and unlocking queues, as well as managing the queue pool, is what explains this bad performance.
    \item \textbf{NVMe Async, Thread Queues:} As expected, this implementation is the fastest for most of the scale factors. It shows that having a queue for each thread,
    that does not need to be managed centrally, is the best implementation, and even improves the most upon the existing DuckDB implementation.
\end{itemize}

\section{Conclusion}

We modified DuckDB to bypass the POSIX file system and directly leverage xNVMe asynchronous I/Os.
Our report shows that only the solution based on thread-owned NVMe queues passes our sanity check.
As expected, NVMe passthrough provides a significant performance improvement, but it is definitely not orders of magnitudes.
This study illustrates the potential and drawbacks of vertical integration of the storage stack with a data-intensive systemo.
On the one hand, the design of DuckDB leads to data dependencies that require careful handling when introducing asynchronous I/Os.
On the other hand, the design of DuckDB is a great match for (i) directly mapping DuckDB blocks onto the LBA space and (ii)
leveraging thread owned NVMe queues.

Directions for future work include performance evaluation beyond sanity check: Can the integration we propose lead to performance improvements?
Is it possible to leverage the DuckDB design to efficiently combine raw device and file system, or more interestingly raw devices used as caches
and object stores? How to conduct DuckDB-SSD co-design? Would DuckDB benefit from specialized Parquet SSDs?

\bibliographystyle{ACM-Reference-Format}
\bibliography{main}


\begin{thebibliography}{20}


\ifx \showCODEN    \undefined \def \showCODEN     #1{\unskip}     \fi
\ifx \showISBNx    \undefined \def \showISBNx     #1{\unskip}     \fi
\ifx \showISBNxiii \undefined \def \showISBNxiii  #1{\unskip}     \fi
\ifx \showISSN     \undefined \def \showISSN      #1{\unskip}     \fi
\ifx \showLCCN     \undefined \def \showLCCN      #1{\unskip}     \fi
\ifx \shownote     \undefined \def \shownote      #1{#1}          \fi
\ifx \showarticletitle \undefined \def \showarticletitle #1{#1}   \fi
\ifx \showURL      \undefined \def \showURL       {\relax}        \fi
\providecommand\bibfield[2]{#2}
\providecommand\bibinfo[2]{#2}
\providecommand\natexlab[1]{#1}
\providecommand\showeprint[2][]{arXiv:#2}

\bibitem[Haas et~al\mbox{.}(2020)]%
        {Haas2020-rj}
\bibfield{author}{\bibinfo{person}{Gabriel Haas}, \bibinfo{person}{Michael
  Haubenschild}, {and} \bibinfo{person}{Viktor Leis}.}
  \bibinfo{year}{2020}\natexlab{}.
\newblock \showarticletitle{{Exploiting directly-attached NVMe arrays in
  {DBMS}}}. In \bibinfo{booktitle}{\emph{{CIDR}}}.
\newblock
\urldef\tempurl%
\url{https://www.cidrdb.org/cidr2020/papers/p16-haas-cidr20.pdf}
\showURL{%
\tempurl}


\bibitem[Haas and Leis(2023)]%
        {Haas2023-lw}
\bibfield{author}{\bibinfo{person}{Gabriel Haas} {and} \bibinfo{person}{Viktor
  Leis}.} \bibinfo{year}{2023}\natexlab{}.
\newblock \showarticletitle{{What Modern NVMe Storage Can Do, and How to
  Exploit it: High-Performance I/O for High-Performance Storage Engines}}.
\newblock \bibinfo{journal}{\emph{Proceedings VLDB Endowment}}
  \bibinfo{volume}{16}, \bibinfo{number}{9} (\bibinfo{date}{May}
  \bibinfo{year}{2023}), \bibinfo{pages}{2090--2102}.
\newblock
\showISSN{2150-8097}
\href{https://doi.org/10.14778/3598581.3598584}{doi:\nolinkurl{10.14778/3598581.3598584}}


\bibitem[Hall and Bonnet(2005)]%
        {Hall2005-db}
\bibfield{author}{\bibinfo{person}{Christoffer Hall} {and}
  \bibinfo{person}{Philippe Bonnet}.} \bibinfo{year}{2005}\natexlab{}.
\newblock \showarticletitle{{Getting priorities straight: improving Linux
  support for database I/O}}. In \bibinfo{booktitle}{\emph{{Proceedings of the
  31st international conference on Very large data bases}}}
  \emph{(\bibinfo{series}{VLDB '05})}. \bibinfo{publisher}{VLDB Endowment},
  \bibinfo{pages}{1116--1127}.
\newblock
\showISBNx{9781595931542}
\href{https://doi.org/10.5555/1083592.1083721}{doi:\nolinkurl{10.5555/1083592.1083721}}


\bibitem[Holanda(2022)]%
        {duckdb-art_blog}
\bibfield{author}{\bibinfo{person}{Pedro Holanda}.}
  \bibinfo{year}{2022}\natexlab{}.
\newblock \bibinfo{title}{Persistent Storage of Adaptive Radix Trees in
  DuckDB}.
\newblock
\newblock
\shownote{Last accessed:
  2025-05-12\\\url{https://duckdb.org/2022/07/27/art-storage.html}}.


\bibitem[Houlborg et~al\mbox{.}(2026)]%
        {cidr26}
\bibfield{author}{\bibinfo{person}{Emil Houlborg},
  \bibinfo{person}{Andreas~Nicolaj Tietgen}, \bibinfo{person}{Simon A.~F.
  Lund}, \bibinfo{person}{Marcel Weisgut}, \bibinfo{person}{Tilmann Rable},
  \bibinfo{person}{Javier Gonzalez}, \bibinfo{person}{Vivek Shah}, {and}
  \bibinfo{person}{Pinar Tozun}.} \bibinfo{year}{2026}\natexlab{}.
\newblock \showarticletitle{{Flexible I/O for Database Management Systems with
  xNVMe}} \emph{(\bibinfo{series}{CIDR})}.
\newblock


\bibitem[Joshi et~al\mbox{.}(2024)]%
        {Joshi2024-nd}
\bibfield{author}{\bibinfo{person}{Kanchan Joshi}, \bibinfo{person}{Anuj
  Gupta}, \bibinfo{person}{Javier Gonz´lez}, \bibinfo{person}{Ankit Kumar},
  \bibinfo{person}{Krishna~Kanth Reddy}, \bibinfo{person}{Arun George},
  \bibinfo{person}{Simon Lund}, {and} \bibinfo{person}{Jens Axboe}.}
  \bibinfo{year}{2024}\natexlab{}.
\newblock \showarticletitle{{I/O Passthru: upstreaming a flexible and efficient
  I/O path in Linux}}. In \bibinfo{booktitle}{\emph{{Proceedings of the 22nd
  USENIX Conference on File and Storage Technologies}}}
  \emph{(\bibinfo{series}{FAST '24}, \bibinfo{number}{Article 7})}.
  \bibinfo{publisher}{USENIX Association}, \bibinfo{address}{USA},
  \bibinfo{pages}{107--122}.
\newblock
\showISBNx{9781939133380}
\href{https://doi.org/10.5555/3650697.3650704}{doi:\nolinkurl{10.5555/3650697.3650704}}


\bibitem[Kuiper and Mühleisen(2023)]%
        {Kuiper2023-pz}
\bibfield{author}{\bibinfo{person}{Laurens Kuiper} {and}
  \bibinfo{person}{Hannes Mühleisen}.} \bibinfo{year}{2023}\natexlab{}.
\newblock \showarticletitle{{These Rows Are Made for Sorting and That’s Just
  What We’ll Do}}. In \bibinfo{booktitle}{\emph{{2023 IEEE 39th International
  Conference on Data Engineering (ICDE)}}}. \bibinfo{publisher}{IEEE},
  \bibinfo{pages}{2050--2062}.
\newblock
\showISSN{2375-026X,1063-6382}
\href{https://doi.org/10.1109/ICDE55515.2023.00159}{doi:\nolinkurl{10.1109/ICDE55515.2023.00159}}


\bibitem[Kuiper et~al\mbox{.}(2021)]%
        {Kuiper2021-uy}
\bibfield{author}{\bibinfo{person}{Laurens Kuiper}, \bibinfo{person}{Mark
  Raasveldt}, {and} \bibinfo{person}{H Mühleisen}.}
  \bibinfo{year}{2021}\natexlab{}.
\newblock \showarticletitle{{Efficient External Sorting in {DuckDB}}}.
\newblock \bibinfo{journal}{\emph{BICOD}} (\bibinfo{year}{2021}),
  \bibinfo{pages}{40--45}.
\newblock
\urldef\tempurl%
\url{https://ceur-ws.org/Vol-3163/BICOD21_paper_9.pdf}
\showURL{%
\tempurl}


\bibitem[Kuschewski et~al\mbox{.}(2024)]%
        {Kuschewski2024-ty}
\bibfield{author}{\bibinfo{person}{Maximilian Kuschewski},
  \bibinfo{person}{Jana Giceva}, \bibinfo{person}{Thomas Neumann}, {and}
  \bibinfo{person}{Viktor Leis}.} \bibinfo{year}{2024}\natexlab{}.
\newblock \showarticletitle{{High-performance query processing with NVMe
  arrays: Spilling without killing performance}}.
\newblock \bibinfo{journal}{\emph{Proc. ACM Manag. Data}} \bibinfo{volume}{2},
  \bibinfo{number}{6} (\bibinfo{date}{Dec.} \bibinfo{year}{2024}),
  \bibinfo{pages}{1--27}.
\newblock
\showISSN{2836-6573}
\href{https://doi.org/10.1145/3698813}{doi:\nolinkurl{10.1145/3698813}}


\bibitem[Leis et~al\mbox{.}(2013)]%
        {duckdb-art_paper}
\bibfield{author}{\bibinfo{person}{Viktor Leis}, \bibinfo{person}{Alfons
  Kemper}, {and} \bibinfo{person}{Thomas Neumann}.}
  \bibinfo{year}{2013}\natexlab{}.
\newblock \showarticletitle{The adaptive radix tree: ARTful indexing for
  main-memory databases}. In \bibinfo{booktitle}{\emph{2013 IEEE 29th
  International Conference on Data Engineering (ICDE)}}. IEEE,
  \bibinfo{pages}{38--49}.
\newblock


\bibitem[Lund(2020)]%
        {Lund2020-hj}
\bibfield{author}{\bibinfo{person}{Simon Lund}.}
  \bibinfo{year}{2020}\natexlab{}.
\newblock \bibinfo{title}{{{xNVMe: Programming Emerging Storage Interfaces for
  Productivity and Performance}}}.
\newblock
  \bibinfo{howpublished}{\url{https://www.snia.org/educational-library/xnvme-programming-emerging-storage-interfaces-productivity-and-performance-2020}}.
\newblock


\bibitem[Lund and Shah(2024)]%
        {xnvme-codesign}
\bibfield{author}{\bibinfo{person}{Simon~AF Lund} {and} \bibinfo{person}{Vivek
  Shah}.} \bibinfo{year}{2024}\natexlab{}.
\newblock \showarticletitle{xNVMe: Unleashing Storage Hardware-Software
  Co-design}.
\newblock \bibinfo{journal}{\emph{arXiv preprint arXiv:2411.06980}}
  (\bibinfo{year}{2024}).
\newblock


\bibitem[Lund et~al\mbox{.}(2022a)]%
        {Lund2022-sx}
\bibfield{author}{\bibinfo{person}{Simon A~F Lund}, \bibinfo{person}{Philippe
  Bonnet}, \bibinfo{person}{Klaus B~A Jensen}, {and} \bibinfo{person}{Javier
  Gonzalez}.} \bibinfo{year}{2022}\natexlab{a}.
\newblock \showarticletitle{{I/O interface independence with {xNVMe}}}. In
  \bibinfo{booktitle}{\emph{{Proceedings of the 15th {ACM} International
  Conference on Systems and Storage}}} \emph{(\bibinfo{series}{SYSTOR '22})}.
  \bibinfo{publisher}{Association for Computing Machinery},
  \bibinfo{address}{New York, NY, USA}, \bibinfo{pages}{108–119}.
\newblock
\showISBNx{9781450393805}
\href{https://doi.org/10.1145/3534056.3534936}{doi:\nolinkurl{10.1145/3534056.3534936}}


\bibitem[Lund et~al\mbox{.}(2022b)]%
        {xnvme-presentation}
\bibfield{author}{\bibinfo{person}{Simon A.~F. Lund}, \bibinfo{person}{Philippe
  Bonnet}, \bibinfo{person}{Klaus B.~A. Jensen}, {and} \bibinfo{person}{Javier
  Gonzalez}.} \bibinfo{year}{2022}\natexlab{b}.
\newblock \showarticletitle{I/O interface independence with xNVMe}. In
  \bibinfo{booktitle}{\emph{Proceedings of the 15th ACM International
  Conference on Systems and Storage}} (Haifa, Israel)
  \emph{(\bibinfo{series}{SYSTOR '22})}. \bibinfo{publisher}{Association for
  Computing Machinery}, \bibinfo{address}{New York, NY, USA},
  \bibinfo{pages}{108–119}.
\newblock
\showISBNx{9781450393805}
\href{https://doi.org/10.1145/3534056.3534936}{doi:\nolinkurl{10.1145/3534056.3534936}}


\bibitem[Mühleisen(2024)]%
        {djikstra}
\bibfield{author}{\bibinfo{person}{Hannes Mühleisen}.}
  \bibinfo{year}{2024}\natexlab{}.
\newblock \bibinfo{title}{Leaving The Two Tier Architecture Behind}.
\newblock \bibinfo{howpublished}{Presented at the CWI Lectures \& Dijkstra
  Fellowship for Marcin Żukowski}.
\newblock
\urldef\tempurl%
\url{https://www.youtube.com/watch?v=H1N2Jr34jwU&pp=ygUoTGVhdmluZyBUaGUgVHdvIFRpZXIgQXJjaGl0ZWN0dXJlIEJlaGluZA%3D%3D}
\showURL{%
\tempurl}


\bibitem[Neumann and Freitag(2020)]%
        {Neumann2020-it}
\bibfield{author}{\bibinfo{person}{Thomas Neumann} {and}
  \bibinfo{person}{Michael~J Freitag}.} \bibinfo{year}{2020}\natexlab{}.
\newblock \showarticletitle{{Umbra: {A} Disk-Based System with In-Memory
  Performance}}. In \bibinfo{booktitle}{\emph{{10th Conference on Innovative
  Data Systems Research, {CIDR} 2020, Amsterdam, The Netherlands, January
  12-15, 2020, Online Proceedings}}}. \bibinfo{publisher}{www.cidrdb.org}.
\newblock
\urldef\tempurl%
\url{http://cidrdb.org/cidr2020/papers/p29-neumann-cidr20.pdf}
\showURL{%
\tempurl}


\bibitem[Raasveldt and M{\"u}hleisen(2019)]%
        {duckdb-design_details}
\bibfield{author}{\bibinfo{person}{Mark Raasveldt} {and}
  \bibinfo{person}{Hannes M{\"u}hleisen}.} \bibinfo{year}{2019}\natexlab{}.
\newblock \showarticletitle{Duckdb: an embeddable analytical database}. In
  \bibinfo{booktitle}{\emph{Proceedings of the 2019 international conference on
  management of data}}. \bibinfo{pages}{1981--1984}.
\newblock


\bibitem[Raasveldt and M{\"u}hleisen(2020)]%
        {duckdb-embedded_olap}
\bibfield{author}{\bibinfo{person}{Mark Raasveldt} {and}
  \bibinfo{person}{Hannes M{\"u}hleisen}.} \bibinfo{year}{2020}\natexlab{}.
\newblock \showarticletitle{Data Management for Data Science-Towards Embedded
  Analytics.}. In \bibinfo{booktitle}{\emph{CIDR}}.
\newblock


\bibitem[Raasveldt and Mühleisen(2019)]%
        {Raasveldt2019-lp}
\bibfield{author}{\bibinfo{person}{Mark Raasveldt} {and}
  \bibinfo{person}{Hannes Mühleisen}.} \bibinfo{year}{2019}\natexlab{}.
\newblock \showarticletitle{{DuckDB: an Embeddable Analytical Database}}. In
  \bibinfo{booktitle}{\emph{{Proceedings of the 2019 International Conference
  on Management of Data}}}. \bibinfo{publisher}{ACM}, \bibinfo{address}{New
  York, NY, USA}.
\newblock
\showISBNx{9781450356435}
\href{https://doi.org/10.1145/3299869.3320212}{doi:\nolinkurl{10.1145/3299869.3320212}}


\bibitem[von Merzljak et~al\mbox{.}(2022)]%
        {Von_Merzljak2022-fu}
\bibfield{author}{\bibinfo{person}{Leonard von Merzljak},
  \bibinfo{person}{Philipp Fent}, \bibinfo{person}{Thomas Neumann}, {and}
  \bibinfo{person}{Jana Giceva}.} \bibinfo{year}{2022}\natexlab{}.
\newblock \showarticletitle{{What are you waiting for? Use coroutines for
  asynchronous I/O to hide I/O latencies and maximize the read bandwidth!}}
\newblock  (\bibinfo{year}{2022}), \bibinfo{pages}{36--46}.
\newblock
\urldef\tempurl%
\url{https://github.com/L-v-M/async}
\showURL{%
\tempurl}


\end{thebibliography}

\end{document}